\newcommand{\ba}{\begin{array}}
\newcommand{\ea}{\end{array}}
\newcommand{\bc}{\begin{center}}
\newcommand{\ec}{\end{center}}

\newcommand{\beqn}[1]{\begin{equation}\label{#1}}
\newcommand{\eeqn}{\end{equation}}
\newcommand{\be}{\begin{equation}}
\newcommand{\ee}{\end{equation}}

\newcommand{\beqnn}{\begin{eqnarray}}
\newcommand{\eeqnn}{\end{eqnarray}}

\newcommand{\col}{{\rm col}}

\newcommand{\diag}{{\rm diag}}

\newcommand{\T}{{\rm T}}

\documentclass[twocolumn]{autart}    

\usepackage{graphicx}          
\usepackage{amsmath}
\usepackage{graphicx}
\usepackage{amssymb}
\usepackage{cite}
\usepackage{cases}
\usepackage{eurosym}
\usepackage{booktabs}
\usepackage{nccmath}
\usepackage[caption=false]{subfig}
\usepackage{bm}
\usepackage{epstopdf}
\usepackage{algorithm}  
\usepackage{algpseudocode}
\usepackage{amssymb}
\usepackage{mathtools}

\usepackage[dvipsnames]{xcolor}
\definecolor{red1}{cmyk}{0, 0.7808, 0.4429, 0.1412}
\usepackage{hyperref}
\hypersetup{
  colorlinks, linkcolor=red
}
\usepackage{romannum}

\begin{document}
\begin{frontmatter}
\title{Resilient and constrained consensus against adversarial attacks: A distributed MPC framework\thanksref{footnoteinfo}}

\thanks[footnoteinfo]{This paper was not presented at any IFAC 
meeting.}

\author[uvic]{Henglai Wei}\ead{henglaiwei@uvic.ca},    
\author[uvic]{Kunwu Zhang}\ead{kunwu@uvic.ca},   
\author[buaa]{Hui Zhang}\ead{huizhang285@gmail.com},             
\author[uvic]{Yang Shi\thanksref{footnoteinfo2}}\ead{yshi@uvic.ca}  

\thanks[footnoteinfo2]{Corresponding author}
\address[uvic]{Department of Mechanical Engineering, University of Victoria, Victoria, BC, Canada, V8W 3P6}  
\address[buaa]{School of Transportation Science
and Engineering, Beihang University, Beijing, China, 100191}             

\begin{keyword}                           
Distributed MPC, Resilient consensus, Constrained consensus, MAS, Adversarial attacks              
\end{keyword}                             
\begin{abstract}                          
There has been a growing interest in realizing the resilient consensus of the multi-agent system (MAS) under cyber-attacks, which aims to achieve the consensus of normal agents (i.e., agents without attacks) in a network, depending on the neighboring information. The literature has developed mean-subsequence-reduced (MSR) algorithms for the MAS with $F$ adversarial attacks and has shown that the consensus is achieved for the normal agents when the communication network is at least ($2F+1$)-robust. However, such a stringent requirement on the communication network needs to be relaxed to enable more practical applications. Our objective is, for the first time, to achieve less stringent conditions on the network, while ensuring the \emph{resilient consensus} for the general linear MAS subject to \emph{control input constraints}. In this work, we propose a distributed resilient consensus framework, consisting of a pre-designed consensus protocol and distributed model predictive control (DMPC) optimization, which can help significantly reduce the requirement on the network robustness and effectively handle the general linear constrained MAS under adversarial attacks. By employing a novel distributed adversarial attack detection mechanism based on the history information broadcast by neighbors and a convex set (i.e., \emph{resilience set}), we can evaluate the reliability of communication links. Moreover, we show that the recursive feasibility of the associated DMPC optimization problem can be guaranteed. The proposed consensus protocol features the following properties: 1) by minimizing a group of control variables, the consensus performance is optimized; 2) the resilient consensus of the general linear constrained MAS subject to $F$-locally adversarial attacks is achieved when the communication network is ($F+1$)-robust. Finally, numerical simulation results are presented to verify the theoretical results.
\end{abstract}

\end{frontmatter}

\section{Introduction}
In the last two decades, much effort has been devoted to investigating consensus problems of the multi-agent system (MAS), which requires agents to achieve an agreement of common interest. In such problems, the information is exchanged in a distributed fashion without the central collection and process. Although many decent consensus protocols have been developed for the MAS (e.g., \cite{olfati2004consensus,ren2005consensus}), there are two fundamental and challenging issues in practice. Conventional consensus research works only consider the normal communication networks among the agents without cyber-attacks. However, the potential malicious intrusions and adversarial attacks may exist in the networks, leading to the systems vulnerability or even damage. Resilience becomes critical when some malicious agents in the network do not obey the pre-defined communication rule and try to mislead the other agents \cite{ishii2022overview}. Therefore, the first issue is, besides the consensus convergence analysis, how to guarantee the resilience of the MAS against adversarial attacks. On the other hand, most of the existing consensus algorithms are not directly applicable for the MAS with state and control input constraints. The second issue is to develop a novel consensus protocol that can ensure the satisfaction of constraints and optimize the consensus performance.

It is well known that distributed model predictive control (DMPC) emerges as an attractive solution for the MAS due to its distinct advantages in tackling various physical constraints, making predictions, and optimizing the control performance. However, existing DMPC-based consensus algorithms do not consider the resilience of the MAS, which greatly motivates us to address these two challenging issues by designing a novel DMPC-based resilient consensus algorithm for the MAS subject to adversarial attacks and constraints.

We now briefly review related works on \emph{DMPC-based consensus} and \emph{resilient consensus}.

\emph{DMPC-based consensus:} Most of the existing results of DMPC focus on the formation stabilization problems of the MAS \cite{dunbar2006distributed,wang2017distributed}. Recent advances in DMPC algorithms lead to a profound interest in consensus problems, and previous works along this research line include the MAS with integrator dynamics \cite{ferrari2009model,zhan2013consensus,li2015neighbor}, the general linear MAS \cite{li2015receding,hirche2020distributed,wang2020distributed,wang2021linear}, and the nonlinear MAS \cite{muller2012cooperative}. In \cite{ferrari2009model}, an MPC-based consensus protocol is proposed for the MAS with single- and double-integrator dynamics over the time-varying networks, where the geometric properties of the optimal path are used to prove the consensus convergence. These two papers \cite{li2015neighbor,li2015receding} investigate the consensus problem of the first-order and general linear MAS and develop explicit consensus protocols via solving unconstrained DMPC problems; the necessary and sufficient conditions are derived to ensure the consensus. In \cite{wang2020distributed}, the optimal consensus problem of the asynchronous MAS with single- and double-integrator dynamics is addressed via a DMPC algorithm, in which the control inputs and the final consensus states are regarded as decision variables. The authors further extend this method to the general linear MAS without considering the physical constraints in \cite{wang2021linear}, in which a consensus manifold is introduced such that the final consensus state and inputs are treated as the augmented decision variables of the DMPC optimization problem. It is worth noting that these DMPC-based consensus algorithms do not investigate the resilience issue of the constrained MAS under the adversarial attacks.

\emph{Resilient consensus:} For consensus problems with security issues, early resilient consensus can be found in one of the seminal papers \cite{lamport1982byzantine}, where different subsystems of the MAS can reach agreement in the presence of Byzantine attacks. Recently, some notable resilient consensus algorithms are reported to alleviate the effects of attacks and ensure the security of the MAS in \cite{pasqualetti2012consensus,leblanc2013resilient,dibaji2017resilient,ruan2019secure,dibaji2018resilient,fiore2019resilient,senejohnny2019resilience,mustafa2020resilient,cetinkaya2020randomized,shang2020resilient,usevitch2020resilient,rezaee2021resiliency,wehbe2021probabilistic,zhao2022resilient,fang2021secure}. If attacks occur, the \emph{detection and identification} for misbehaving agents becomes crucial from the security viewpoint. Early methods in \cite{pasqualetti2012consensus,sundaram2011distributed} adopt a bank of unknown-input observers to detect and identify the fault behaviors or malicious attacks. Note that these methods require global knowledge of the communication topology and sufficient computational resources. To enable fully distributed implementation, Byzantine-resilient distributed observers are developed in \cite{mitra2019byzantine}. The Mean-Subsequence-Reduced (MSR) algorithm developed in \cite{leblanc2013resilient} enables the MAS with integrator dynamics to tolerate a finite number $F$ of adversaries and still to reach resilient consensus when the communication graph satisfies certain robustness properties. More specifically, the MSR algorithm discards neighbors' suspicious or extreme scalar state values and updates the control input with the remaining normal ones. There are also several modified MSR-type algorithms for resilient consensus in the literature \cite{dibaji2017resilient,dibaji2018resilient,usevitch2020resilient,fiore2019resilient,shang2020resilient}; however, it turns out to be technically challenging to guarantee that the resulting states stay in the convex hull of normal system states for the higher-dimensional MAS in $\mathbb{R}^n$, $n\geq2$. The authors in \cite{zhao2017resilient,zhao2022resilient} resort to the distributed attack isolation algorithm to realize resilient consensus by excluding the attacked agents. In most of these methods, the resilient consensus of the MAS in the presence of $F$ adversarial attacks can be achieved if the communication network is at least $(2F+1)$-robust. It is worth mentioning that the above-mentioned resilient consensus methods for the MAS are studied in the absence of constraints. \emph{Given these facts, a question naturally arises: Is it possible to develop a detection and identification algorithm that relaxes the restriction on the network robustness (i.e., ($2F+1$)-robust) while guaranteeing the resilient consensus of the general linear constrained MAS in the presence of $F$ adversarial attacks?}

\emph{Contributions:} In this work, we give the affirmative answer to this question. To the best of our knowledge, this is the first work of tackling the resilient consensus problem of the general linear MAS with input constraints.
\begin{itemize}
\item First, a DMPC-based resilient consensus framework is proposed for the general linear MAS with input constraints. Note that most of the existing resilient consensus algorithms are only applicable for the MAS with the simple integrator dynamics, not to mention the case where the constraints are taken into account. Moreover, the proposed DMPC-based consensus algorithm is implemented in parallel since the previous optimal predicted trajectories are employed to estimate the current optimal predicted trajectories with the bounded estimation error guarantee.
\item Second, in contrast to existing resilient consensus works \cite{pasqualetti2012consensus,leblanc2013resilient,dibaji2017resilient,ruan2019secure,dibaji2018resilient,fiore2019resilient,senejohnny2019resilience,mustafa2020resilient,cetinkaya2020randomized,shang2020resilient,usevitch2020resilient,rezaee2021resiliency,wehbe2021probabilistic,zhao2022resilient,fang2021secure}, where each agent discards $F$ largest and $F$ smallest state values received from its neighbors, in this paper, we propose a novel distributed detection algorithm that only requires the normal agent to detect and discard at most $F$ adversarial attacks. The proposed algorithm greatly reduces the requirement on the robustness of the communication networks, which constitutes one of our main contributions. 
\end{itemize}
The rest of this paper is organized as follows: Section \ref{whl9-sec:2} provides the adversarial attack model and formulates the resilient consensus problem. In Section \ref{whl9-sec:3}, we present the estimation error set and the resilient set. The DMPC-based resilient consensus framework is developed for the constrained MAS in Section \ref{whl9-sec:4}. Section \ref{whl9-sec:5} gives the theoretical analysis of the recursive feasibility and the resilient consensus convergence. Section \ref{whl9-sec:6} gives the simulation results of the proposed method before the conclusion in Section \ref{whl9-sec:7}.

\emph{Notations}: The symbols $\mathbb{N}_{\geq 0}$ and $\mathbb{N}_{[m,n]}$ denote the sets of the nonnegative integers and integers in the interval $[m,n]$, respectively. For ${x}\in \mathbb{R}^n$, $\|{x}\|$ denotes the Euclidean norm, $\|{x}\|_{{P}}^2$ denotes the weighted Euclidean norm ${x}^\text{T}P{x}$, where $P$ is positive definite. $[{x}_1^\text{T},\dots,{x}_n^\text{T}]^\text{T}$ is written as $\col({x}_1,\dots,{x}_n)$. Given two sets $\mathcal{X},\mathcal{Y}\subseteq \mathbb{R}^n$, the set operation $\mathcal{X}\backslash \mathcal{Y}$ is defined as $\mathcal{X}\backslash \mathcal{Y}:=\{{x}\mid {x}\in\mathcal{X},{x}\notin \mathcal{Y}\}$. The set addition is $\mathcal{X}\oplus\mathcal{Y}:=\{x+y\mid x\in\mathcal{X},y\in\mathcal{Y}\}$, and the set subtraction is $\mathcal{X}\ominus\mathcal{Y}:=\{x\in\mathbb{R}^n\mid x\oplus\mathcal{Y}\subseteq\mathcal{X}\}$. $I_M\in\mathbb{R}^{M\times M}$ is the identity matrix, and $\bm{1}$ is the all-ones vector with proper dimensions. $\bar{\lambda}({Q})$ and $\underline{\lambda}(Q)$ denote the largest and smallest eigenvalues of the matrix $Q$, respectively. $\rho(Q)$ denotes the spectral radius of the matrix $Q$. The symbol $\otimes$ denotes the Kronecker product. ${x}(t)$ denotes the state ${x}$ at time $t$, and ${x}(t+k|t)$ denotes the predicted state within the controller at some future time $t+k$ determined at time $t$.

\section{Preliminaries and problem formulation}
\label{whl9-sec:2}
\subsection{Graph theory}
Consider a group of agents interacting over a time-dependent undirected graph $\mathcal{G}(t)=\{\mathcal{V},\mathcal{E}(t)\}$, where $\mathcal{V}=\{1,2,\dots,M\}$ is the set of agents, and the edge set $\mathcal{E}(t)\subseteq\mathcal{V}\times \mathcal{V}$ describes the time-varying connections between agents, with $t\in\mathbb{N}_{\geq 0}$. An edge $(i,j)\in\mathcal{E}(t)$ implies that agent $i$ can exchange information with agent $j$ at time $t$. Let $\mathcal{A}(t)=[a_{ij}(t)]\in\mathbb{R}^{M\times M}$ be the weighted adjacency matrix of $\mathcal{G}(t)$, with the $(i,j)$-entry $a_{ij}(t)>\alpha$ and $0<\alpha<1$. Agent $i$ and agent $j$ are called broadcaster and receiver, respectively, when agent $i$ broadcasts the information to the neighbor $j$. The set of agent $i$'s neighbors at time $t$ is denoted by $\mathcal{N}_i(t)=\{j\in\mathcal{V}\mid(i,j)\in\mathcal{E}(t),i\neq j\}$. The weighted Laplacian matrix $\mathcal{L}(t)\in\mathbb{R}^{M\times M}$ is symmetric with $\mathcal{L}_{ii}(t)=\sum_{j\in \mathcal{N}_i(t)}a_{ij}(t)\leq 1$, $\mathcal{L}_{ij}(t)=-a_{ij}(t)$. The cardinality of a set $\mathcal{N}$ is denoted as $|\mathcal{N}|$. 

The following $r$-robustness notations introduced in \cite{leblanc2013resilient} guarantee the connectivity of a graph $\mathcal{G}$ when some agents discard a certain number of communication links. 

\begin{defn}($r$-reachable set \cite{leblanc2013resilient})
Given a graph $\mathcal{G}$ and a nonempty subset $\mathcal{S}\subset\mathcal{V}$, the set $\mathcal{S}$ is $r$-reachable if $\exists\ i\in\mathcal{S}$ such that $|\mathcal{N}_i\backslash \mathcal{S}|\geq r$, $r\in\mathbb{N}_{\geq 1}$.
\end{defn}

\begin{defn}($r$-robust graph \cite{leblanc2013resilient})
A nonempty graph $\mathcal{G}$ is $r$-robust ($r<M$) if for any pair of nonempty disjoint subsets of $\mathcal{V}$, at least one of the subsets is $r$-reachable.
\end{defn}

Note that a time-dependent graph $\mathcal{G}(t)$ is $r$-robust if for its temporal topology is an $r$-robust graph.

\subsection{Resilient consensus of the constrained MAS}
Consider a group of agents, and agent $i$, $i\in \mathcal{V}$ is described by
\begin{equation}\label{whl9-eq:1}
x_i(t+1)=Ax_i(t)+Bu_i(t), \ t\in\mathbb{N}_{\geq 0},
\end{equation} 
where $x_i(t)\in\mathbb{R}^n$ and $u_i(t)\in\mathbb{R}^m$ are the system state and control input of agent $i$, respectively. Each agent satisfies the control input constraint set, i.e., 
\begin{equation}\label{whl9-eq:2}
u_i(t)\in \mathcal{U}_i\subset\mathbb{R}^m, 
\end{equation}
where $\mathcal{U}_i$ contains the origin. 


In what follows, we introduce the notations of adversarial links/agents and resilient consensus \cite{leblanc2013resilient}. Let $\mathcal{V}_N(t)\subset\mathcal{V}$ and $\mathcal{V}_A(t)\subset\mathcal{V}$ denote the set of normal and adversarial agents at time $t$, respectively. Analogously, the edge set can be partitioned into two disjoint subsets: a set of the normal links $\mathcal{E}_N(t)$ and a set of adversarial links $\mathcal{E}_A(t)$. 

\begin{defn}(Normal agent) An agent in \eqref{whl9-eq:1} is normal if it updates system states and broadcasts information based on the designed consensus protocol.
\end{defn}

\begin{defn}(Adversarial agent)
An edge $(i,j)\in\mathcal{E}(t)$ is adversarial if it transmits arbitrary value from agent $i$ to agent $j$, $i\in\mathcal{V}$, $j\in\mathcal{N}_i(t)$. An agent in \eqref{whl9-eq:1} is adversarial if it broadcasts arbitrarily different state values to its neighbors. 
\end{defn}

Note that the definition of adversarial agents in this work covers both the malicious agent and Byzantine agent. The malicious agent sends the same misbehaving information to all of its neighbors; the Byzantine agent sends the different misbehaving information to its neighbors. The adversarial behaviors in this paper are categorized into two types: $1)$ Adversarial links are removed from the graph network, in which only the broadcast information to its neighbors is malicious; $2)$ Adversarial agents are removed from the original graph network, where the adversarial agents include the malicious and Byzantine attackers. 

\begin{defn}($F$-locally adversarial graph)
The graph $\mathcal{G}$ is $F$-locally adversarial, if for each normal agent $i$, $i\in\mathcal{V}_N(t)$, the number of adversarial neighboring links/agents is no more than $F$, i.e., $|\mathcal{E}_A\cap \{(j,i),j\in\mathcal{N}_i\}|\leq F$ or $|\mathcal{N}_i\cap \mathcal{V}_A|\leq F$.
\end{defn}

It is assumed that an upper bound for the number of adversarial links/agents $F$ is available for each normal agent. For simplicity, we call the adversarial links/agents as adversarial attacks hereafter. Let $\{t_1^i,t_2^i\dots,t_F^i\}$ denote the attack time sequence of agent $i$, $i\in\mathcal{V}_A(t)$ with $t_\tau^i\in\mathbb{N}_{\geq 1}$ and $\tau\in\mathbb{N}_{[1,F]}$.

Our objective in this paper is to design a distributed resilient consensus protocol such that the constrained MAS with $F$-locally adversarial attacks over the $F+1$ robust communication network that attains
\begin{itemize}
\item[1)] \emph{Resilient agreement}: For the MAS \eqref{whl9-eq:1} in the presence of $F$-locally adversarial attacks, it holds that $\lim_{t\to\infty}\|x_i(t)-x_j(t)\|=0$, $i,j\in\mathcal{V}_N(t)$, $j\in\mathcal{N}_i(t)$.
\item[2)] \emph{Constraint satisfaction}: Agent $i$, $i\in\mathcal{V}_N(t)$
\begin{equation*}
x_i(t+1)=Ax_i(t)+Bu_i(t), 
\end{equation*}
satisfies the constraint \eqref{whl9-eq:2} for all $t\in\mathbb{N}_{\geq 0}$.
\end{itemize}

\section{Estimation error set and resilience set}
\label{whl9-sec:3}
This section specifies an estimation error set for the broadcaster to achieve the parallel implementation of the distributed control algorithm requiring only neighbor-to-neighbor communication. Furthermore, a resilience set is designed for the receiver, based on which we develop the distributed attack detection algorithm.

\subsection{Estimation error set}
\label{whl9-sec:3-1}
Most of the existing DMPC algorithms assume that the communication among agents is perfect, and the information can be exchanged simultaneously (e.g., see \cite{wang2017distributed,li2018receding}). However, it is unfeasible for agents to calculate the predicted state sequences while exchanging them simultaneously in practice. Alternatively, the assumed predicted state sequences are exchanged among the MAS in this paper. Exactly, the assumed predicted state sequence (i.e., the predicted state sequence broadcast at the previous time instant $t$) is used to estimate the optimal predicted state sequence at $t+1$, $t\in\mathbb{N}_{\geq 0}$. Let $\hat{\bm{x}}_i(t+1)=\col(\hat{x}_i(t+1|t+1),\dots, \hat{x}_i(t+1+N|t+1))$ be the assumed state sequence of agent $i$, $i\in\mathcal{V}$ hereafter; $\hat{x}_i(t+1+k|t+1)$ is constructed as
\begin{equation}\label{whl9-eq:3}
\hat{x}_i(t+k|t+1)=x_i^*(t+k|t), k\in \mathbb{N}_{[1,N+1]}, 
\end{equation}
in which $t\in\mathbb{N}_{\geq 0}$, and $u_i^*(\cdot|t)$ and $x_i^*(\cdot|t)$ are, respectively, the optimal control inputs and the optimal predicted states generated by solving the DMPC optimization problem $\mathcal{P}_i$ defined in Section \ref{whl9-sec:4} at time $t$. 

In the case of the parallel implementation, the estimation errors induced by the non-simultaneous transmission can be regarded as external disturbances. Then a set $\Delta$ is designed to restrict the estimation error such that the deviation between the actual predicted state and the broadcast assumed predicted state is bounded. That is, the actual predicted state of agent $i$, $i\in\mathcal{V}$ is required to lie in a neighborhood of the assumed predicted state
\begin{equation}\label{whl9-eq:4}
{x}_i(t+k|t)\in \hat{x}_i(t+k|t)\oplus \Delta,\ k\in\mathbb{N}_{\geq 0},
\end{equation}
with $\Delta=\{\delta\in\mathbb{R}^n\mid\|\delta\|\leq \eta,\eta>0\}$.

\subsection{Resilience set}
In this work, the communication graph $\mathcal{G}$ is assumed to have $F$-locally adversarial attacks. Note that each normal agent has to detect and discard adversarial attacks to eliminate the potential adverse effect; otherwise, the consensus of the MAS cannot be reached. Intuitively, each agent should have more than $F$ neighbors to guarantee the connectivity of the graph $\mathcal{G}$ with adversarial attacks. Here, a resilience set $\mathcal{R}$ is designed for the receiver based on the estimation error set to detect and identify adversarial attacks, which is defined by $\mathcal{R}=\Delta$. Next, the receiver can categorize the received predicted state information of neighbors into two types as follows. 
\begin{itemize}
\item[1)] \emph{Normal communication:} If the predicted state of neighbors $\hat{{x}}_j(t+k|t)$, $j\in\mathcal{N}_i(t)$ satisfies
\begin{equation}\label{whl9-eq:5}
\hat{x}_j(t+k|t)-\hat{x}_j(t+k|t-1)\in\mathcal{R},\ k\in\mathbb{N}_{\geq 0},
\end{equation}
then the link $(i,j)$ is normal. From \eqref{whl9-eq:5}, it is observed that the assumed predicted state $\hat{x}_j(t+k|t-1)$ broadcast at time $t-1$ serves as the center of the tube $R_j(t)$. 
\item[2)] \emph{Adversarial communication:} If the received assumed predicted state sequence $\hat{x}_j(t+k|t)$ satisfies
\begin{equation}\label{whl9-eq:6}
\hat{x}_j(t+k|t)-\hat{x}_j(t+k|t-1)\notin\mathcal{R},\ k\in\mathbb{N}_{\geq 0},
\end{equation}
then the communication between agent $i$ and $j$ is adversarial. The adversarial predicted state sequence will be discarded and not involved in the consensus protocol design for agent $i$ (which implies $a_{ij}(t)=0$).
\end{itemize}



Note that only the adversarial predicted state sequence induced by the cyber-attacks needs to be detected based on the broadcast assumed predicted state sequence and the resilience set as in \eqref{whl9-eq:5} and \eqref{whl9-eq:6}. Notice that we assume that no adversarial attacks occur in the MAS at the initial time instant.
\begin{rem}
Incorporating the constraint \eqref{whl9-eq:4} into the DMPC optimization problem can restrain the deviation between the intended behavior of agent $i$ and what its neighbor $j$, $j\in\mathcal{N}_i(t)$ believes how agent $i$ will behave. In this way, the distributed optimization problems are solved in parallel with the pre-specified bound on the estimation errors. Similar constraint is also studied in \cite{dunbar2006distributed,liu2018distributed}. Note that this constraint is considered only from the \emph{broadcaster} perspective. However, when it comes to the MAS under cyber-attacks, the consistency constraint for the broadcaster in \cite{dunbar2006distributed,liu2018distributed} might not guarantee the resilient consensus since the attackers tamper with the broadcast information in the communication channel. As a distinct feature, the resilient set designed from the \emph{receiver} side aims to detect and identify the adversarial attacks, enabling the resilient consensus of the MAS to be achieved. 
\end{rem}
\subsection{Distributed detection algorithm}

According to the conditions in \eqref{whl9-eq:5} and \eqref{whl9-eq:6}, the broadcast information is discarded once the adversarial attacks are detected. The distributed attack detection algorithm is designed for each agent as follows.
\begin{algorithm}[H]
\caption{Distributed attack detection algorithm}\label{whl9-alg:1}
\textbf{Input:} The broadcast predicted state sequence $\hat{\bm{x}}_j(t)$ of agent $j$, $j\in\mathcal{N}_i(t-1)$, $t\in\mathbb{N}_{\geq 1}$.\\
\textbf{Output:} The neighboring set $\mathcal{N}_i(t)$ and the weight $a_{ij}(t)$.
\begin{algorithmic}[1]
\Require The estimation error set $\Delta$, the assumed predicted states $\hat{\bm{x}}_j(t-1)$ and $a_{ij}(t-1)$, with $a_{ij}(0)=1/|\mathcal{N}_i(0)|$.
\For{$j=1$ to $|\mathcal{N}_i(t-1)|$}
\State Receive the assumed predicted states $\hat{\bm{x}}_j(t)$;
\If {The condition \eqref{whl9-eq:6} holds}
    \State $a_{ij}(t)\gets 0$;
    \State $|\mathcal{N}_i(t)|\gets |\mathcal{N}_i(t-1)|-1$;
    \Else 
    \State $a_{ij}(t)\gets a_{ij}(t-1)$;
\EndIf
\EndFor
\end{algorithmic}
\end{algorithm}

\begin{figure}[!ht]
\centering
\subfloat[MSR algorithms]{%
\centering
  \includegraphics[clip,width=0.39\columnwidth]{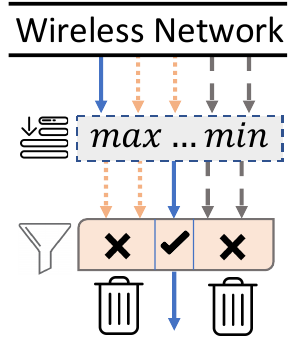}\label{whl9-fig:1-1}%
}
\hspace{18pt}
\subfloat[\textbf{Algorithm \ref{whl9-alg:1}}]{%
\centering
  \includegraphics[clip,width=0.42\columnwidth]{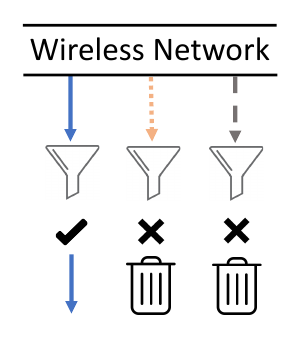}\label{whl9-fig:1-2}%
}
\caption{Illustration of two types of attack detection algorithms for the MAS in the presence of $F=2$ locally adversarial attacks. (a) MSR-type algorithm: Each normal agent requires at least five communication connections from its neighbors. (b) \textbf{Algorithm \ref{whl9-alg:1}}: Each normal agent only requires three communication links from its neighbors.}\label{whl9-fig:2}
\end{figure}

\begin{rem}
The redundant communication links are crucial to achieving resilient consensus for the MAS in the presence of adversarial attacks. MSR-type algorithms have been extensively studied for the resilient consensus problem of the MAS \cite{leblanc2013resilient,dibaji2017resilient,dibaji2018resilient,usevitch2020resilient,fiore2019resilient,shang2020resilient}. As shown in Fig.~\ref{whl9-fig:1-1}, these algorithms usually require that each normal agent gathers all neighbors' information, sorts the received information, and discards $F$ small and $F$ large extreme state values. Intuitively, MSR-type algorithms collect and detect adversarial attacks in a centralized way, resulting in a high requirement for network robustness. To avoid the influence of $F$ adversarial attacks, the communication graph $\mathcal{G}$ has to be $2F+1$ robust. In contrast, the proposed detection algorithm checks the broadcast information from neighbors in a distributed fashion, significantly relaxing the requirement on the network robustness. Notably, each normal agent only ignores at most $F$ neighbors' information for the MAS in the presence of $F$-locally adversarial attacks, as shown in Fig.~\ref{whl9-fig:1-2}. The relaxed robustness of the communication graph constitutes a distinct contribution to this paper.
\end{rem}

\section{DMPC for resilient consensus}
\label{whl9-sec:4}
The proposed DMPC-based consensus protocol consists of the sum of the pre-designed consensus protocol and the online DMPC input.

The DMPC-based resilient consensus scheme for an MAS example of three agents in the presence of adversarial attacks is illustrated in Fig.~\ref{whl9-fig:2}. For each agent $i$, $i\in\mathcal{V}$, the scheme mainly consists of six parts: the controlled system, DMPC controller, broadcaster, receiver, attack detector and wireless network. \textbf{Algorithms \ref{whl9-alg:1}} is performed in the attack detector.
\begin{figure}[!ht]
\centering
\includegraphics[width=0.95\columnwidth]{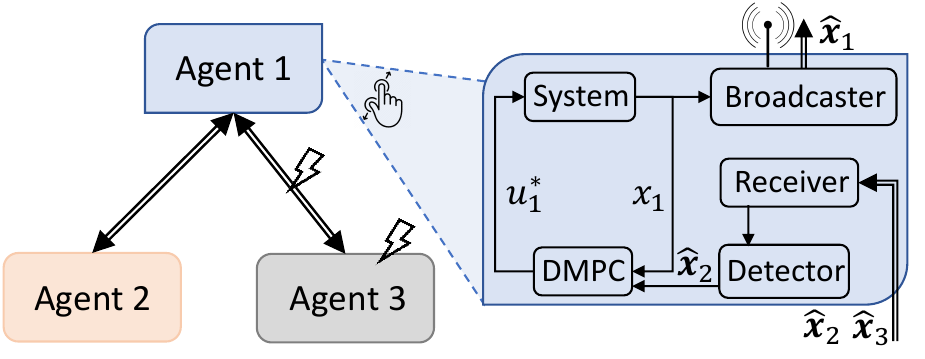}
\caption{The DMPC-based resilient consensus scheme for the MASs (consisting of three agents with $\mathcal{N}_1=\{2,3\}$, $\mathcal{N}_2=\{1\}$ and $\mathcal{N}_3=\{1\}$). In the wireless communication network, the information between agents is transmitted using broadcast mechanism.}
\label{whl9-fig:2}
\end{figure}

\subsection{Pre-designed consensus protocol}
At time $t$, the control input for agent $i$ is determined based on the relative states with its neighbors, i.e.,
\begin{equation}\label{whl9-eq:9}
u_i(t)=K(t)\sum_{j\in\mathcal{N}_i(t)}a_{ij}(t)(x_i(t)-x_j(t))+c_i(t),
\end{equation}
where $\varkappa_i(t)=K(t)\sum_{j\in\mathcal{N}_i(t)}a_{ij}(t)(x_i(t)-x_j(t))$ is the pre-designed consensus input, and $c_i(t)$ is a decision variable of the DMPC optimization problem $\mathcal{P}_i$. Note that the adversarial attacks may yield a time-dependent communication network. 

The ``pre-stabilizing'' control for the stabilization problem of the single system in \cite{chisci2001systems} is extended to consider the consensus problem of the constrained MAS. The advantages of the proposed protocol lie in the following aspects: 1) The pre-designed protocol $\varkappa_i(t)$ achieves consensus for the unconstrained MAS concerning a quadratic performance criterion \cite{movric2014cooperative}. $c_i(t)$ is calculated online via solving the DMPC optimization problem. The proposed consensus protocol can provide a suboptimal consensus performance while ensuring the satisfaction of the constraints. 2) Compared with the existing DMPC methods \cite{zhan2018distributed,li2018receding}, the knowledge of the communication topology is explicitly exploited to construct the pre-designed consensus protocol, which facilitates the consensus convergence analysis of the constrained MAS.

\subsection{DMPC for constrained consensus}
Each agent has a local cost function $J_i(\bm{c}_i(t))$, 
\begin{equation}
J_i(\bm{c}_i(t))=\sum_{k=0}^{N}\|{c}_i(t+k|t)\|^2_{\Psi},
\end{equation}
where the weighting matrix $\Psi$ is positive definite, $N\in\mathbb{N}_{\geq 1}$ is the prediction horizon, and $\bm{c}_i(t)=\col(c_i(t|t),\dots,c_i(t+N|t))$ is the control sequence. At time $t$, given the system state $x_i(t)$ of agent $i$, $i\in \mathcal{V}$ and its neighbors' assumed predicted state sequence $\hat{\bm{x}}_j(t)$, $j\in\mathcal{N}_i(t)$, the DMPC problem $\mathcal{P}_i$ is given as follows
\begin{fleqn}
\begin{subequations}\label{whl9-eq:11}
\begin{alignat}{2}
\min_{\bm{c}_i(t)}\ & {J}_{i}(\bm{c}_i(t)) \notag\\
\text{s.t.}\ & x_i(t|t)=x_i(t),\\
&x_i(t+k+1|t)=Ax_i(t+k|t)+Bu_i(t+k|t),\hskip -0.5em\\
&u_i(t+k|t)=\hat{\varkappa}_i(t+k|t)+c_i(t+k|t),\\
&{{u}}_i(t+k|t)\in {\mathcal{U}}_i,\label{whl9-eq:11d}\\
&x_i(t+k|t)\in\hat{x}_i(t+k|t)\oplus \Delta, \label{whl9-eq:11e}
\end{alignat}
\end{subequations}
\end{fleqn}
where $\hat{\varkappa}_i(t+k|t)=K(t)\sum_{j\in\mathcal{N}_i(t)}a_{ij}(t)(x_i(t+k|t)-\hat{x}_j(t+k|t))$, $k\in\mathbb{N}_{[0,N]}$. Let $\bm{c}_i^*(t)=\col(c_i^*(t|t),\dots,c_i^*(t+N|t))$ be the optimal solution to the optimization problem $\mathcal{P}_i$ at time $t$. We have the optimal control input $u_i^*(\cdot|t)$ for agent $i$
\begin{equation}\label{whl9-eq:12}
\begin{aligned}
&{u}_i^*(t+k|t)\\
=&K(t)\sum_{j\in\mathcal{N}_i(t)}a_{ij}(t)(x_i^*(t+k|t)\\
&-\hat{x}_j(t+k|t))+c_i^*(t+k|t),
\end{aligned}
\end{equation}
with $k\in\mathbb{N}_{\geq 0}$. The optimal control input sequence at $t$ is $\bm{u}_i^*(t)=\col({u}_i^*(t|t),\dots,u_i^*(t+N|t))$ and the corresponding optimal predicted state is
\begin{equation}
x_i^*(t+k+1|t)=Ax_i^*(t+k|t)+Bu_i^*(t+k|t),
\end{equation}
where the optimal state sequence is denoted by $\bm{x}_i^*(t)=\col(x_i^*(t|t),\dots,x_i^*(t+N|t))$. Further, applying the first term of the optimal control input in \eqref{whl9-eq:12} to the system in \eqref{whl9-eq:1} yields the closed-loop system immediately
\begin{equation}
x_i(t+1)=Ax_i(t)+Bu_i^*(t|t).
\end{equation}

\begin{assum}\label{whl9-asm:1}
For agent $i$, $i\in\mathcal{V}$, there exists a sufficient large prediction horizon $N$ for the DMPC optimization problem $\mathcal{P}_i$ such that $\hat{\varkappa}_i(t+k|t)\in\mathcal{U}_i$, $k\in\mathbb{N}_{\geq N}$.
\end{assum}

Let $\zeta_i(t+k|t)=x_i(t+k|t)-\bar{x}_{-i}(t+k|t)$ be the average predicted error state, where $\bar{x}_{-i}(t+k|t)=\sum_{j\in\mathcal{N}_i(t)}a_{ij}(t)x_j(t+k|t)$. The original consensus problem $\mathcal{P}_i$ can be converted into a stabilization problem. Given the initial state $\zeta_i(t|t)$ and the optimal control input $u_i^*(t+k|t)$, the techniques for the stabilizing MPC in \cite{boccia2014stability} can be used to find a finite prediction horizon $N$ fulfilling \textbf{Assumption \ref{whl9-asm:1}}.

Note that each agent can verify the information from its neighbors at each time step based on \textbf{Algorithm \ref{whl9-alg:1}} to detect and isolate the adversarial attacks. The pre-designed consensus gain $K(t)=-1/\lambda_M(t) B^\T \Psi A/(B^\T \Psi B+R)$ is recalculated once attacks occur, where $\lambda_M(t)$ is the maximum eigenvalue of the graph $\mathcal{G}(t)$ \cite{you2011network}. $\lambda_M(t)$ is updated when the graph topology $\mathcal{G}(t)$ changes. The overall DMPC-based consensus algorithm is summarized in \textbf{Algorithm \ref{whl9-alg:2}}. 

\begin{algorithm}[H]
\caption{DMPC-based consensus protocol}\label{whl9-alg:2}
\begin{algorithmic}[1]
\Require The weighting matrix $\Psi$, the set $\Delta$, the initial assumed state $\hat{x}_i(k|0)=A^kx_i(0)$, $k\in\mathbb{N}_{\geq 0}$, the pre-designed feedback gain $K(0)$, and other parameters. Set ${t}=1$. \label{step1} 
\While{For agent $i$, the control is not stopped}
\State Measure the current system state $x_i(t)$; 
\State \parbox[t]{210pt}{Receive and evaluate the assumed state sequence of neighbors $\hat{\bm{x}}_j(t)$, $j\in \mathcal{N}_i(t)$ as in \textbf{Algorithm}~\ref{whl9-alg:1};}\label{step2}
\If{$a_{ij}(t)==0$}
\State Update the gain $K(t)$;
\Else
\State $K(t)=K(t-1)$;
\EndIf
\State \parbox[t]{210pt}{Solve the optimization problem $\mathcal{P}_i$ in \eqref{whl9-eq:11} to generate the optimal control sequence $\bm{c}_i^*(t)$ and $\bm{u}_{i}^{*}(t)$, and the optimal predicted state sequence $\bm{x}_i^*(t)$;}\label{step3}  
\State Apply the control input ${u}_i^*(t|t)$ to agent $i$; 
\State \parbox[t]{210pt}{Broadcast the assumed predicted sequence $\hat{x}_i(t+k|t)$, $t\in\mathbb{N}_{\geq 0}$ as in \eqref{whl9-eq:3} to agent $j$, $j\in\mathcal{N}_i(t)$;}
\State ${t}={t}+1$;
\EndWhile\label{endwhile}
\end{algorithmic}
\end{algorithm}

\begin{rem}
We note that most of the existing results on resilient consensus have only dealt with the unconstrained MAS with single-integrator dynamics (see, e.g., \cite{leblanc2013resilient,dibaji2017resilient,dibaji2018resilient,fiore2019resilient,senejohnny2019resilience,usevitch2020determining}), which cannot be directly applied for the constrained MAS with general linear dynamics. Thus, the proposed DMPC-based resilient consensus algorithm for the constrained MAS is more general and practical. 
\end{rem}

Compared with the conventional DMPC algorithms, the computational complexity of the proposed algorithm does not increase when the MAS is attacked since the attacked agents will be isolated and removed from the communication networks, which releases the corresponding computational resources. Meanwhile, only the gain $K(t)$ needs to be recalculated. As a result, the proposed algorithm can be applied for the large-scale MAS with suitable robust networks.

\section{Theoretical analysis}
\label{whl9-sec:5}
\begin{figure}[!ht]
\includegraphics[width=0.9\columnwidth]{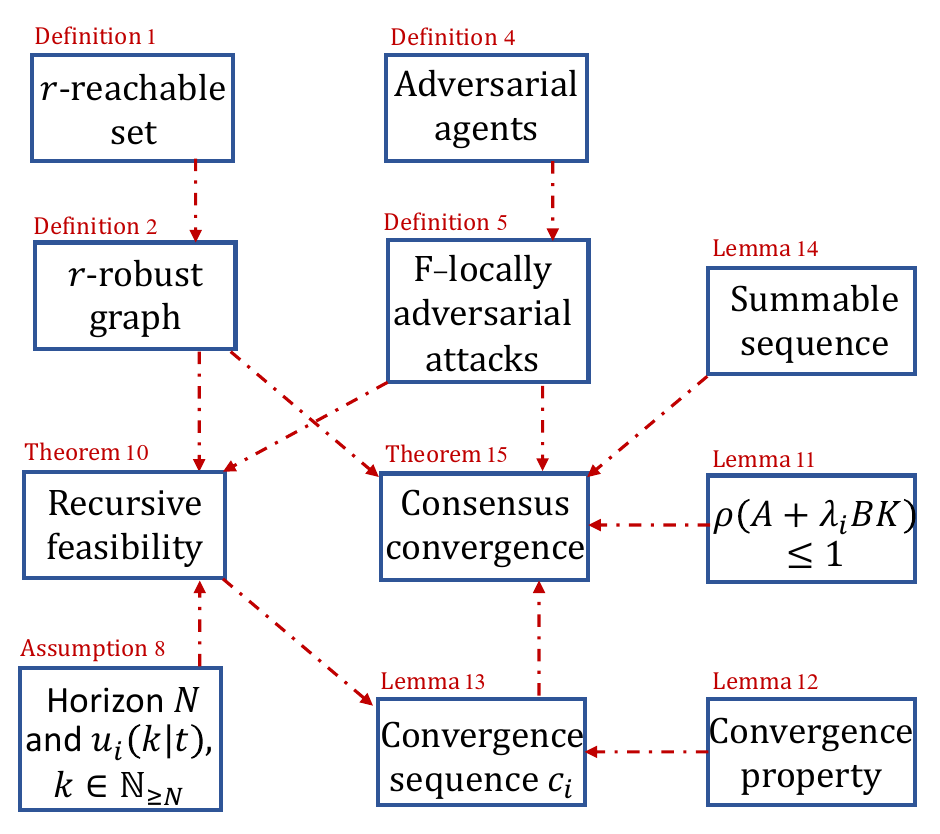}
\centering
\caption{Illustration of the relation among the recursive feasibility and the resilient consensus convergence.}
\label{whl9-fig:3}
\end{figure}

Before proceeding to the theoretical analysis, we illustrate the relationship of the theoretical results, as shown in Fig. \ref{whl9-fig:3}.

\subsection{Recursive feasibility analysis}

The adversarial attacks on the MAS may result in a varying communication network and a different consensus gain $K(t)$. Consequently, we discuss the feasibility of the cases with and without attacks in the sequel.

\begin{thm}\label{whl9-thm:1}
For the constrained MAS with $F$-locally adversarial attacks, suppose that \textbf{Assumption \ref{whl9-asm:1}} is satisfied. If the conditions 
\begin{equation}\label{whl9-eq:13a}
\rho(\sum_{s=0}^{N}A_K^{N-s}BK(t))\leq 1,
\end{equation}
and
\begin{equation}\label{whl9-eq:14a}
\rho(A_K)<1
\end{equation}
are satisfied, with $A_K=A+BK(t)$, the adversarial attacks can be detected via \textbf{Algorithm \ref{whl9-alg:1}}, and the optimization problem $\mathcal{P}_i$ in \eqref{whl9-eq:11} is feasible at $t$, $i\in\mathcal{V}_N(t)$, $t\in\mathbb{N}_{\geq 0}$, then it admits a feasible solution at $t+1$.
\end{thm}
\begin{pf}
Suppose that the optimal solution to the local optimization problem in \eqref{whl9-eq:11} exists at time $t$. We first demonstrate that there are no attacks on the MAS (\textbf{Case 1}), then show that there exist attacks (\textbf{Case 2}). 

\noindent\textbf{Case 1: No attacks occur at $t+1$:}
A candidate input sequence $\tilde{\bm{c}}_i(t+1)$ at $t+1$ is then created by dropping the first input and appending a terminal zero element of the optimal control at $t$, that is,
\begin{equation}\label{whl9-eq:15}
\begin{aligned}
&\tilde{c}_i(t+1+k|t+1)=c_i^*(t+1+k|t), \ k\in\mathbb{N}_{[0,N-1]},\\
&\tilde{c}_i(t+1+N|t+1)=0.
\end{aligned}
\end{equation}
Note that the assumed predicted state sequence $\hat{\bm{x}}_j(t+1)$, $j\in\mathcal{N}_i(t+1)$, can be received by agent $i$ at $t+1$, the control input sequence $\tilde{\bm{u}}_i(t+1)=\col(\tilde{u}_i(t+1|t+1),\dots,\tilde{u}_i(t+1+N|t+1))$ for agent $i$ is constructed as
\begin{equation}
\begin{aligned}
&\tilde{u}_i(t+1+k|t+1)\\
=&K(t)\sum_{j\in\mathcal{N}_i}a_{ij}(t)(\tilde{x}_i(t+1+k|t+1)\\
&-\hat{x}_j(t+1+k|t+1))+\tilde{c}_i(t+1+k|t+1),\label{whl9-eq:16}
\end{aligned}
\end{equation}
where $K(t+1)= K(t)$, $a_{ij}(t+1)=a_{ij}(t)$, $k\in\mathbb{N}_{\geq 0}$ and the system state $\tilde{x}_i(t+1+k|t+1)$ satisfies the following difference equation
\begin{equation}\label{whl9-eq:17}
\begin{aligned}
&\tilde{x}_i(t+1+k+1|t+1)\\
=&A\tilde{x}_i(t+1+k|t+1)+B\tilde{u}_i(t+1+k|t+1),
\end{aligned}
\end{equation}
with the initial condition $\tilde{x}_i(t+1|t+1)=x_i(t+1)$. By \eqref{whl9-eq:12} and \eqref{whl9-eq:17}, we know that $x_i(t+1)=x_i^*(t+1|t)$.

From \eqref{whl9-eq:15} and \eqref{whl9-eq:16}, one gets 
\begin{equation*}
\begin{aligned}
&\tilde{u}_i(t+1|t+1)\\
=&K(t)\sum_{j\in\mathcal{N}_i}a_{ij}(t)\big(\tilde{x}_i(t+1|t+1)-\hat{x}_j(t+1|t)\\
&+\hat{x}_j(t+1|t)-\hat{x}_j(t+1|t+1)\big)+\tilde{c}_i(t+1|t+1)\\
=&K(t)\sum_{j\in\mathcal{N}_i}a_{ij}(t)\big(x_i^*(t+1|t)-\hat{x}_j(t+1|t)\big)+c_i^*(t+1|t)\\
&+K(t)\sum_{j\in\mathcal{N}_i}a_{ij}(t)\big(\hat{x}_j(t+1|t)-x_j^*(t+1|t)\big)\\
=&u_i^*(t+1|t)+K(t)\sum_{j\in\mathcal{N}_i}a_{ij}(t)\big(\hat{x}_j(t+1|t)-{x}_j^*(t+1|t))\big)\\
=&u_i^*(t+1|t)+K(t)w_i(t+1|t).
\end{aligned}
\end{equation*}

Indeed, substituting the control input $\tilde{u}_i(t+1|t+1)$ to the system in \eqref{whl9-eq:17} yields 
\begin{equation}
\begin{aligned}
&\tilde{x}_i(t+2|t+1)\\
=&A\tilde{x}_i(t+1|t+1)+B\tilde{u}_i(t+1|t+1)\\
=&Ax_i^*(t+1|t)+Bu_i^*(t+1|t)+BK(t)w_i(t+1|t)\\
=&x_i^*(t+2|t)+BK(t)w_i(t+1|t).
\end{aligned}
\end{equation}
Proceeding forward, the state sequence candidate is then derived as 
\begin{equation}\label{whl9-eq:19}
\begin{aligned}
&\tilde{x}_i(t+1+k|t+1)\\
=&x_i^*(t+1+k|t)\\
&+\sum_{s=0}^{k-1}A_K^{k-1-s}BK(t)w_i(t+1+s|t).
\end{aligned}
\end{equation}

Define the variable $r_i^k(t+1)\in\mathbb{R}^n$ for agent $i$ by
\begin{equation*}
\begin{aligned}
r_i^k(t+1)=
\begin{dcases}
0, & k=0,\\
\sum_{s=0}^{k-1}A_K^{k-1-s}BK(t)w_i(t+s|t),&  k\in\mathbb{N}_{[1,N]},
\end{dcases}
\end{aligned}
\end{equation*}
and one obtains the corresponding set 
\begin{equation}\label{whl9-eq:21}
\begin{aligned}
\mathcal{R}_i^k(t+1)=\begin{dcases}
0, & k=0,\\
\bigoplus_{s=0}^{k-1}A_K^{k-1-s}BK(t)\mathcal{W}_i,& t\in\mathbb{N}_{[1,N]},
\end{dcases}
\end{aligned}
\end{equation}
where $r_i^k(t+1)\in{\mathcal{R}}_i^k(t+1)$.
Since $\sum_{j\in\mathcal{N}_i(t+1)}a_{ij}(t+1)\leq 1$, and $\rho(\sum_{s=0}^{N-1}A_K^{N-1-s}BK(t))\leq 1$, it holds that $\mathcal{R}_i^N=\bigoplus_{s=0}^{N-1}A_K^{N-s}BK(t)\mathcal{W}_i\in\Delta$. Thus, we claim that the constraint \eqref{whl9-eq:11e} of $\mathcal{P}_i$ is satisfied.

Let $\bar{u}_i^*(t+k|t)=K(t)\sum_{j\in\mathcal{N}_i}a_{ij}(t)\big(x_i^*(t+k|t)-x_j^*(t+k|t)\big)+c_i^*(t+k|t)$ be the nominal optimal control input with $k\in\mathbb{N}_{[0,N-1]}$. The nominal optimal control input satisfies $\bar{u}_i^*(t+k|t)\in\overline{\mathcal{U}}_i$, where $\overline{\mathcal{U}}_i=\mathcal{U}_i\ominus K(t)\Delta$. It should be mentioned that the nominal optimal control input $\bar{u}_i^*(\cdot|t)$ is not implemented to the actual system, since $x_j^*(t+k|t)$, $j\in\mathcal{N}_i(t)$ is not available for agent $i$, $i\in\mathcal{V}_N(t)$ at time $t$. But it is used in the analysis of the feasibility. Next, substituting \eqref{whl9-eq:19} into \eqref{whl9-eq:16} yields
\begin{equation}\label{whl9-eq:22}
\tilde{u}_i(t+k|t+1)=\bar{u}_i^*(t+k|t)+K(t)r_i^k(t+1),
\end{equation}
where $k\in\mathbb{N}_{[1,N]}$. Since the optimal control input trajectory $\bm{u}_i^*(t)$ is assumed to be feasible at time $t$, we then have $\bar{u}_i^*(t+k|t)\in\overline{\mathcal{U}}_i$, $k\in\mathbb{N}$. In addition, $r_i^k(t+1)\in\Delta$. Hence, at time $t+1$, from \eqref{whl9-eq:16} and \eqref{whl9-eq:22}, we get
\begin{equation}
\tilde{u}_i(t+k|t+1)\in \mathcal{U}_i,
\end{equation}
with $k\in\mathbb{N}_{[1,N]}$. Furthermore, \textbf{Assumption \ref{whl9-asm:1}} ensures that $\tilde{u}_i(t+1+N|t+1)\in\mathcal{U}_i$. Thus, the control input constraint in \eqref{whl9-eq:11d} holds. Thus, the recursive feasibility is ensured for the case of attack-free at time $t+1$.

\noindent\textbf{Case 2: Attacks occur at $t+1:$} A candidate input sequence $\tilde{\bm{c}}_i(t+1)$ is constructed based on the optimal control sequence $u_i^*(t+k|t)$, i.e.,
\begin{equation}\label{whl9-eq:24}
\begin{aligned}
&\tilde{c}_i(t+1+k|t+1)\\
=&u_i^*(t+1+k|t)-K(t+1)\sum_{j\in\mathcal{N}_i(t+1)}a_{ij}(t+1)\\
&\big(\tilde{x}_i(t+1+k|t+1)-\hat{x}_j(t+1+k|t+1)\big), 
\end{aligned}
\end{equation}
with $k\in\mathbb{N}_{[0,N]}$. The control input sequence $\tilde{\bm{u}}_i(t+1)=\col(\tilde{u}_i(t+1|t+1),\dots,\tilde{u}_i(t+1+N|t+1))$ then becomes
\begin{equation}\label{whl9-eq:25}
\tilde{u}_i(t+1+k|t+1)=u_i^*(t+1+k|t),
\end{equation}
where $k\in\mathbb{N}_{[0,N-1]}$. In addition, one can construct $\tilde{c}_i(t+1+N|t+1)$ such that $\tilde{u}_i(t+1+N|t+1)\in\mathcal{U}_i$. Therefore, the control constraint \eqref{whl9-eq:11d} is ensured at time $t+1$.

With the initial condition $\tilde{x}_i(t+1|t+1)=x_i^*(t+1|t)$, then the corresponding system state $\tilde{x}_i(t+1+k|t+1)$ will be 
\begin{equation}\label{whl9-eq:23}
\tilde{x}_i(t+1+k|t+1)={x}_i^*(t+1+k|t),
\end{equation}
where $k\in\mathbb{N}_{[0,N]}$. Because of $x_i^*(t+N+1|t)=Ax_i^*(t+N|t)+Bu_i^*(t+N|t)$ and $\tilde{u}_i(t+N|t+1)=u_i^*(t+N|t)$, we obtain $\tilde{x}_i(t+1+N|t+1)=x_i^*(t+1+N|t)$. Hence, the constraint \eqref{whl9-eq:11e} holds.

From \eqref{whl9-eq:25} and \eqref{whl9-eq:23}, the feasibility is established at time $t+1$ when attacks occur. Thereby, the proof of the recursive feasibility is completed. $\hfill\blacksquare$
\end{pf}

In \textbf{Theorem \ref{whl9-thm:1}}, the recursive feasibility of the DMPC-based consensus algorithm is provided for the normal agents. Once agent $i$, $i\in\mathcal{V}$ is attacked at time $t_\tau^i$, then we don't analyze the feasibility of the optimization problem $\mathcal{P}_i$ at time $t'$, with $t'>t^i_\tau$ and $t'\in\mathbb{N}_{\geq 1}$.
\subsection{Consensus analysis}
In this subsection, we first present four technical lemmas and then provide the proof of \textbf{Theorem \ref{whl9-thm:2}}. 

The following lemma (see \cite[Theorem 1]{li2009consensus}) provides a consensus condition for the discrete-time MAS. 

\begin{lem} \label{whl9-lem:1}
For the MAS over the network $\mathcal{G}$, the consensus of the MAS is achieved if and only if there exists a consensus gain $K\in\mathbb{R}^{m\times n}$ such that the inequality $\rho(A+\lambda_iBK)<1$ holds, where $\lambda_i$, $i=2,3,\dots,M$, are the nonzero eigenvalues of the Laplacian matrix $\mathcal{L}$.
\end{lem}

The following results on $\{\theta_k\}$, $\{\mu_k\}$ and $\{\sigma_k\}$ are fundamental to the consensus convergence analysis, and the proof can be found in \cite[Lemma 1]{chang2014distributed}.
\begin{lem} \label{whl9-lem:2}
Let $\{\theta_k\}$, $\{\mu_k\}$ and $\{\sigma_k\}$ be non-negative sequences, suppose $\sum_{k=1}^\infty \sigma_k<\infty$ and 
$$\theta_k\leq \theta_{k-1}-\mu_{k-1}+\sigma_{k-1}, \ \forall k\in \mathbb{N}_{\geq1},$$
then the sequence $\{\theta_k\}$ converges and $\sum_{k=1}^\infty \mu_k<\infty$.
\end{lem}

\begin{lem}\label{whl9-lem:3}
For the MAS \eqref{whl9-eq:1} in the presence of $F$-locally adversarial attacks, if the initial state ${x}_i(0)$ is feasible and $\sum_{k=0}^\infty \|c_i(t+k|t)\|_{\Psi}^2< \infty$, $t\in\mathbb{N}_{\geq 0}$, then the sequence $c_i(t)$ satisfies $\lim_{t\to\infty}c_i(t)=0$.
\end{lem}
\begin{pf}
To prove the convergence of $c_i(t)$ as $t\to \infty$, we introduce the following function
\begin{equation*}
V_i(t)=J_i(\bm{c}_i^*(t))=\sum_{k=0}^{N}\|c_i^*(t+k|t)\|_{\Psi}^2.
\end{equation*}

Consider the control input sequence $\tilde{\bm{c}}_i(t+1)$ in \eqref{whl9-eq:15} for the MAS when there exist no attacks at $t+1$, and we get
\begin{equation*}
\begin{aligned}
\tilde{V}_i(t+1)=&\sum_{k=0}^{N}\|\tilde{c}_i(t+1+k|t+1)\|_{\Psi}^2\\
=&V_i(t)-\|c_i^*(t|t)\|_{\Psi}^2.
\end{aligned}
\end{equation*}

It follows from \textbf{Theorem \ref{whl9-thm:1}} that the control input $\tilde{\bm{c}}_i(t+1)$ is a feasible but not necessarily an optimal solution of the problem $\mathcal{P}_i$ at $t+1$. Then, one has
\begin{equation*}
V_i(t+1)\leq \tilde{V}_i(t+1)=V_i(t)-\|c_i^*(t|t)\|_{\Psi}^2.
\end{equation*}

It holds that
\begin{equation}\label{whl9-eq:32}
V_i(t+1)-V_i(t)\leq-\|c_i^*(t|t)\|_{\Psi}^2.
\end{equation}

Note that there are at most $F$-locally adversarial attacks for agent $i$, $i\in\mathcal{V}$, which implies that the control input candidate in \eqref{whl9-eq:24} are adopted no more than $F$ times during the time interval $\mathbb{N}_{[1,t_F^i]}$. {Also note that $\bar{C}_i=\sum_{\tau=1}^{F}\sum_{k=0}^N \|\tilde{c}_i(t_{\tau}^i+k|t_\tau^i)\|_{\Psi}^2<\infty$}, with $\tau\in\mathbb{N}_{[1,F]}$. 

Upon summing up $V_i(t+1)-V_i(t)$ in \eqref{whl9-eq:32} from $t=0$ to $t=k$, we get
\begin{equation}
\begin{aligned}
&\lim_{k\to\infty}\sum_{t=0}^{k}(V_i(t+1)-V_i(t))\\
\leq &\lim_{k\to\infty}V_i(k+1)-V_i(0)+\bar{C}_i\\
\leq&-\lim_{k\to\infty}\sum_{t=0,t\neq t_\tau^i}^k\|c_i^*(t|t)\|_{\Psi}^2+\bar{C}_i,
\end{aligned}
\end{equation}
and $V_i(t)$ as $t\to \infty$, satisfies
\begin{equation*}
0\leq V_i(\infty)\leq V_i(t)-\lim_{k\to\infty}\sum_{t=0}^k\|c_i^*(t|t)\|_{\Psi}^2+\bar{C}_i< \infty,
\end{equation*}
where $V_i(\infty)=\lim_{t\to\infty}V_i(t)$. Using \textbf{Lemma \ref{whl9-lem:2}}, one obtains that $V_i(t)$ converges as $t\to\infty$. One further has $\lim_{t\to\infty}\|c_i^*(t|t)\|_{\Psi}^2=0$,
which implies that $\lim_{t\to\infty}\|c_i(t)\|=0$. By now, we have shown the convergence of the control variable $c_i(t)$. $\hfill\blacksquare$
\end{pf}

Next, we recall a lemma from \cite[Lemma 7]{nedic2010constrained}.
\begin{lem}\label{whl9-lem:4}
{For any given scalar $\beta\in(0,1)$, suppose that the summable sequence $\{\alpha(t)\}$ satisfies $\lim_{t\to\infty}\alpha(t)=0$, then it holds that $\lim_{k\to\infty}\sum_{t=0}^{k}\beta^{k-t}\alpha(t)=0$}.
\end{lem}
The consensus property of the MAS under the DMPC algorithm is reported as follows.
\begin{thm}\label{whl9-thm:2}
Consider the constrained MAS \eqref{whl9-eq:1} in the presence of $F$-locally adversarial attacks. Suppose that the communication network $\mathcal{G}$ is $F+1$ robust, then the normal agents achieve resilient consensus under the recursively feasible \textbf{Algorithm \ref{whl9-alg:2}}, with $i,j\in\mathcal{V}_N(t)$ and $v'=|\mathcal{V}_N(t)|$.
\end{thm}
\begin{pf}
Substitute \eqref{whl9-eq:11} into \eqref{whl9-eq:4}, and we obtain
\begin{equation}\label{whl9-eq:26}
\bm{x}(t+1)=(I_{v'}\otimes A+\mathcal{L}\otimes BK)\bm{x}(t)+(I_M\otimes B)\bm{c}(t),
\end{equation}
in which $\bm{x}(t)=\col(x_1(t),x_2(t),\dots,x_{v'}(t))$, $\mathcal{L}=\mathcal{L}(t)$, $\bm{c}(t)=\col(c_1(t),c_2(t),\dots,c_{v'}(t))$, and $K=K(t)$. Note that corresponding variables and matrices for the MAS in \eqref{whl9-eq:26} have the compatible dimensions concerning the normal agents. 

The average state of the MAS is defined by $\bar{x}(t)={1/{v'}}(\bm{1}^\T\otimes I_n) \bm{x}(t)\in\mathbb{R}^n$, with $\bm{1}$ denotes a compatible vector with all elements to be $1$. Then,
\begin{equation}
\begin{aligned}
\bar{x}(t+1)=& \frac{1}{v'}(\bm{1}^\T \otimes A)\bm{x}(t)+\frac{1}{v'}(\bm{1}^\T \mathcal{L}\otimes BK)\bm{x}(t)\\
&+\frac{1}{v'}(\bm{1}^\T \otimes B)\bm{c}(t)\\
=& \frac{1}{v'}(\bm{1}^\T \otimes A)\bm{x}(t)+B\bar{c}(t)\\
=&  A\bar{x}(t)+B\bar{c}(t),
\end{aligned}
\end{equation}
with $\bar{c}(t)=(\bm{1}^\T\otimes I_n) \bm{c}(t)/v'$. Define $\xi_i(t)=x_i(t)-\bar{x}(t)$ and $\bm{\xi}=\col(\xi_1,\xi_2,\dots,\xi_{v'})$, $i\in\mathcal{V}_N(t)$, then we have
\begin{equation}\label{whl9-eq:35}
\begin{aligned}
\bm{\xi}(t+1)=&(I_{v'}\otimes A+\mathcal{L}\otimes BK)\bm{\xi}(t)\\
&+(I_{v'}\otimes B)\big((I_{v'}-\frac{\bm{1}^\T\bm{1}}{v'})\otimes I_m\big)\bm{c}(t).
\end{aligned}
\end{equation}
There always exists an orthogonal matrix $U=[\bm{1}/\sqrt{v'},U_2,\dots,U_{v'}]\in\mathbb{R}^{{v'}\times {v'}}$  such that the Laplacian matrix is diagonalized, i.e., $U^\T \mathcal{L}U=\diag(0,\lambda_2,\dots,\lambda_{v'})$, where $U_i$, $i\in \mathbb{N}_{[2,v']}$ is an orthogonal eigenvector of $\mathcal{L}$.

Using the property of Kronecker product, one obtains
\begin{equation*}
\begin{aligned}
&(U^\T\otimes I_n)(I_{v'}\otimes A+\mathcal{L}\otimes BK)(U\otimes I_n)\\
=&\diag(A,A+\lambda_2BK,\dots,A+\lambda_{v'}BK).
\end{aligned}
\end{equation*}
Define $\tilde{\bm{\xi}}(t)=\col(\tilde{\xi}_1(t),\tilde{\xi}_2(t),\dots,\tilde{\xi}_{v'})=(U^\T \otimes I_n)\bm{\xi}(t)$, then \eqref{whl9-eq:35} is expressed by
\begin{equation*}
\begin{aligned}
\tilde{\bm{\xi}}(t+1)=&\diag(A,A+\lambda_2BK,\dots,A+\lambda_{v'}BK)\tilde{\bm{\xi}}(t)\\
&+(U^\T \otimes I_n)(I_{v'}\otimes B)\big((I_{v'}-\frac{\bm{1}^\T\bm{1}}{v'})\otimes I_m\big)\bm{c}(t).
\end{aligned}
\end{equation*}
Next, we define the transition matrix $\Phi=\diag(A,A+\lambda_2BK,\dots,A+\lambda_{v'}BK)$ and $\mathcal{B}=(I_{v'}\otimes B)\big((I_{v'}-{\bm{1}^\T\bm{1}/v'})\otimes I_m\big)$, then \eqref{whl9-eq:35} becomes
\begin{equation*}
\tilde{\bm{\xi}}(t+1)=\Phi\tilde{\bm{\xi}}(t)+\mathcal{B}\bm{c}(t),
\end{equation*}
which implies that $\tilde{\bm{\xi}}(t)=\Phi^{t}\tilde{\bm{\xi}}(0)+\sum_{k=0}^{t-1}\Phi^k\mathcal{B}\bm{c}(t-1-k),\ t\in\mathbb{N}_{\geq 1}.$

It is easy to know that $\tilde{\xi}_1(t)=1/\sqrt{v'}(\sum_{i=1}^{v'}\xi_i(t))=0$. Also, due to $\rho(A+\lambda_iBK)<1$, $i\in\mathbb{N}_{[2,v']}$, we obtain the term $\lim_{t\to\infty}\Phi^{t}\tilde{\bm{\xi}}(0)=0$.

{Since $\rho(A+\lambda_iBK)< 1$, one gets $\|(I_{v'}-{\bm{1}^\T\bm{1}/v'})\otimes I_m\|\rho(A+\lambda_iBK)^t< 1$.} In light of this, there always exists a constant $\beta\in(0,1)$, such that
\begin{equation}\label{whl9-eq:39}
\|\big((I_{v'}-\frac{\bm{1}^\T\bm{1}}{v'})\otimes I_m\big)\Phi^t\|\leq\beta^t<1.
\end{equation}

Define $E(t-1-k)=\|I_{v'}\otimes B\|\|\bm{c}(t-1-k)\|$.  
Using the Cauchy-Schwarz inequality and \eqref{whl9-eq:39}, we have
\begin{equation*}
\begin{aligned}
&\lim_{t\to\infty}\sum_{k=0}^{t-1}\Phi^k\mathcal{B}\bm{c}(t-1-k)\\
\leq&\lim_{t\to\infty}\sum_{k=0}^{t-1}\|(I_{v'}-\frac{\bm{1}^\T\bm{1}}{v'})\otimes I_m\Phi^k\|E(t-1-k)\\
\leq&\lim_{t\to\infty}\sum_{k=0}^{t-1}\beta^kE(t-1-k).
\end{aligned}
\end{equation*}

Using \textbf{Lemma}~\ref{whl9-lem:3} and \textbf{Lemma}~\ref{whl9-lem:4}, we get
\begin{equation}
\lim_{t\to\infty}\sum_{k=0}^{t-1}\Phi^k\mathcal{B}\bm{c}(t-1-k)=0,
\end{equation}
implying that $\lim_{t\to\infty} \|x_i(t)-x_j(t)\|=0$, $i,j\in\mathcal{V}_{N}(t)$. Therefore, the constrained MAS under $F$-locally adversarial attacks reaches resilient consensus. The proof is completed. $\hfill\blacksquare$
\end{pf}

For \textbf{Algorithms \ref{whl9-alg:1}, \ref{whl9-alg:2}} and \textbf{Theorem \ref{whl9-thm:1}}, we make the following remarks.\\
\begin{itemize}
\item[1)] \textbf{Comparison with existing resilient consensus algorithms.} Resilient consensus algorithms for the constrained MAS are rarely studied in the literature \cite{ishii2022overview}. A recent result \cite{shang2020resilient} shows that the resilient consensus of the MAS with state constraints can be achieved. The projection-based resilient consensus method is designed based on MSR-type algorithms, where the control input is determined by continuously checking $f_i(x_i)$ and $\dot{f}_i(x_i)$; $f_i(x_i)$ and $\dot{f}_i(x_i)$ specify the state constraint region and the state changing trend, respectively. Similar to the MSR-type algorithms, the resilient algorithm also has a high requirement for the robustness of the communication networks. Especially, this algorithm only considers the single-integrator dynamics and does not apply the general linear MAS with input constraints. In contrast, our proposed DMPC-based resilient consensus algorithm applies to the general linear MAS with stable, semi-stable, and unstable dynamics.
\item[2)] \textbf{Comparison with existing consensus algorithms.} Most of the existing resilient consensus algorithms only discuss the average consensus problems, e.g., \cite{dibaji2018resilient,shang2020resilient,fiore2019resilient,fang2021secure}, which can be regarded as a special case of the leaderless consensus problem addressed in this work with $A=I_n$ and $B=I_m$. In particular, the proposed method can also be extended to solve the leader-following consensus problems of the constrained MAS against adversarial attacks. 
\item[3)] \textbf{Discussion on the resilient consensus convergence.} Conventional DMPC-based formation stabilization methods (e.g., \cite{muller2012cooperative,dunbar2006distributed}) choose the optimal value function as a Lyapunov function to establish the closed-loop stability. However, given the time-varying predicted state trajectories of neighbors and the time-varying networks induced by the adversarial attacks, it is hard to choose such a suitable Lyapunov function to guarantee the resilient consensus convergence. In this work, the pre-designed consensus protocol introduced in the DMPC-based consensus scheme provides a simple method for the constrained MAS to prove the resilient consensus convergence by definition.

\end{itemize}

\section{Simulation}
\label{whl9-sec:6}
In this section, two simulation examples are provided to illustrate the theoretical results of this article.

\textbf{Example 1: The MAS with unstable dynamics.}

\begin{figure}[!ht]
\centering
\subfloat[Adversarial link]{%
\centering
  \includegraphics[clip,width=0.45\columnwidth]{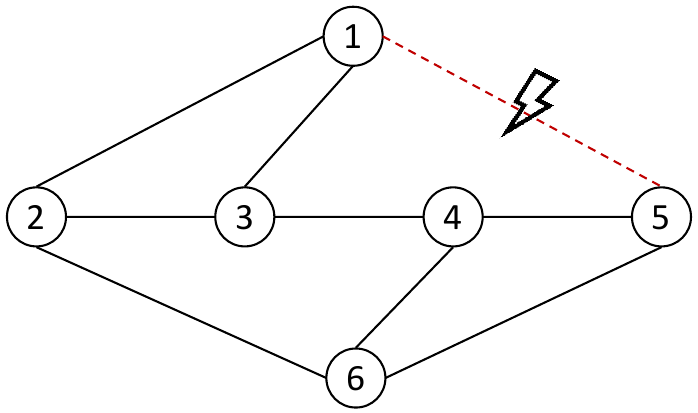}\label{whl9-fig:4-1}%
}
\hfill
\subfloat[Adversarial agent]{%
\centering
  \includegraphics[clip,width=0.45\columnwidth]{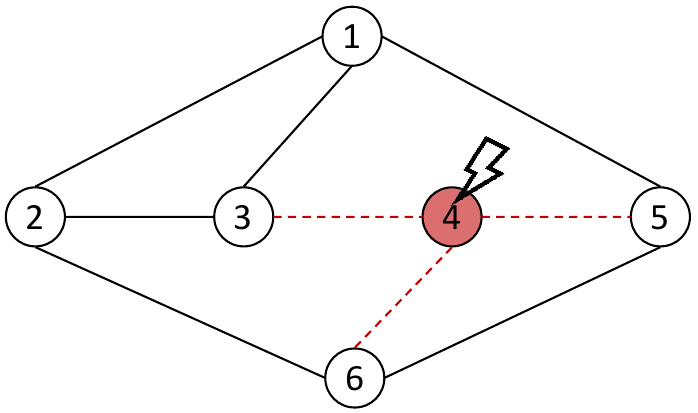}\label{whl9-fig:4-2}%
}
\caption{The $3$-robust graph with six agents under $F=2$ locally adversarial attacks: (a) Adversarial link $(1,5)$, and (b) Adversarial agent $4$.}\label{whl9-fig:4}
\end{figure}

Consider an MAS consisting of six identical discrete-time oscillators, and agent $i$, $i\in\mathcal{V}$ satisfies
\begin{equation}\label{whl9-eq:37}
x_i(k+1)=Ax_i(k)+Bu_i(k), \ i=1,2,\dots,6,
\end{equation}
with $A=[0, 1;-1,0]$ and $B=[0.5;0.5]$.
The control input constraints are $\|u_i\|_{\infty}\leq0.5$. The initial states of six agents are $x_1(0)=[-0.18; 3.21]$, $x_2(0)=[3.32; -1.18]$, $x_3(0)=[-2.29; -2.14]$, $x_4(0)=[-1.22; 2.24]$, $x_5(0)=[1.50; 1.40]$ and $x_6(0)=[-2.42; 0.04]$, respectively. The initial network is configured as $\mathcal{N}_1(0)=\{2,3,5\}$, $\mathcal{N}_2(0)=\{1,3,6\}$, $\mathcal{N}_3(0)=\{1,2,4\}$, $\mathcal{N}_4(0)=\{3,5,6\}$, $\mathcal{N}_5(0)=\{1,4,6\}$ and $\mathcal{N}_6(0)=\{2,4,5\}$. 
The wireless communication network $\mathcal{G}$ is shown in Fig.~\ref{whl9-fig:4}.
The prediction horizon is $N=20$ and the estimation error set is $\Delta=\{x\in\mathbb{R}^2\mid\|x\|\leq0.1\}$. The weighting matrix is ${\Psi}=1$, and the pre-designed consensus feedback matrix is designed as $K(0) = [0.3125,-0.3724]$ with $R=1$. 





\begin{figure}[!ht]
\includegraphics[width=0.85\columnwidth]{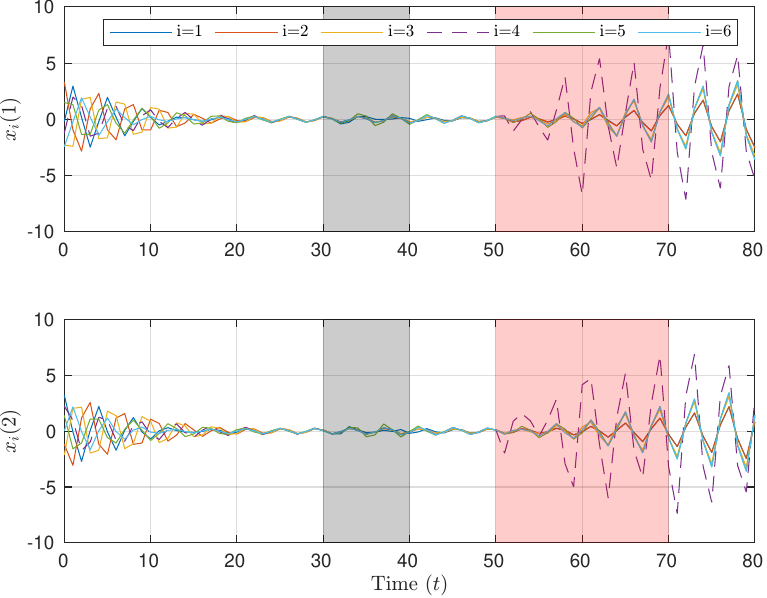}
\centering
\caption{The states $x_i$ of the MAS under $F=2$ locally adversarial attacks without the detection mechanism. The gray and red areas represent the periods under the adversarial link and agent, respectively.}
\label{whl9-fig:5}
\end{figure}

\begin{figure}[!ht]
\includegraphics[width=0.85\columnwidth]{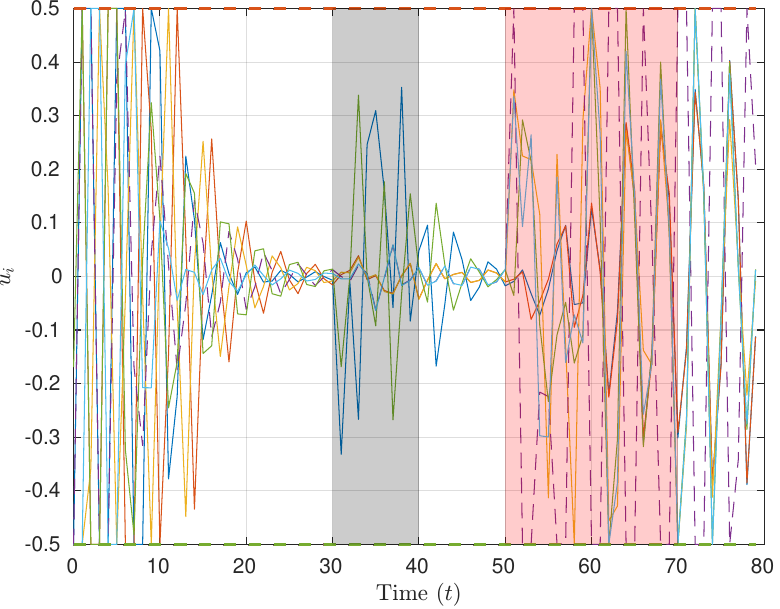}
\centering
\caption{The control inputs $u_i$ of the MAS under $F=2$ locally adversarial attacks without the detection mechanism. The gray and red areas represent the periods under the adversarial link and agent, respectively.}
\label{whl9-fig:6}
\end{figure}
We consider the adversarial agent and link in the simulation. We assume that one out of six agents (i.e., agent $4$) is attacked and becomes the adversarial agent. The agent attack signal $s_a(k)\in\mathbb{R}^2$ is randomly generated from the interval $[-2,2]$ for $k\in\mathbb{N}_{[50,70]}$, and injected into the system in \eqref{whl9-eq:37}. In addition, the link $(1,5)$ is attacked. The adversarial link signal $s_l(k)\in\mathbb{R}^2$ is randomly generated from the interval $[-2,2]$ and incorporated into the broadcast state sequence for $k\in\mathbb{N}_{[30,40]}$. 

When the distributed detection algorithm is not implemented on the MAS, the MAS under $F=2$ locally adversarial attacks, including the adversarial agent and adversarial link, cannot reach consensus as illustrated in Fig. \ref{whl9-fig:5}. The corresponding control inputs of the MAS are given in Fig. \ref{whl9-fig:6}. 

\begin{figure}[!ht]
\includegraphics[width=0.85\columnwidth]{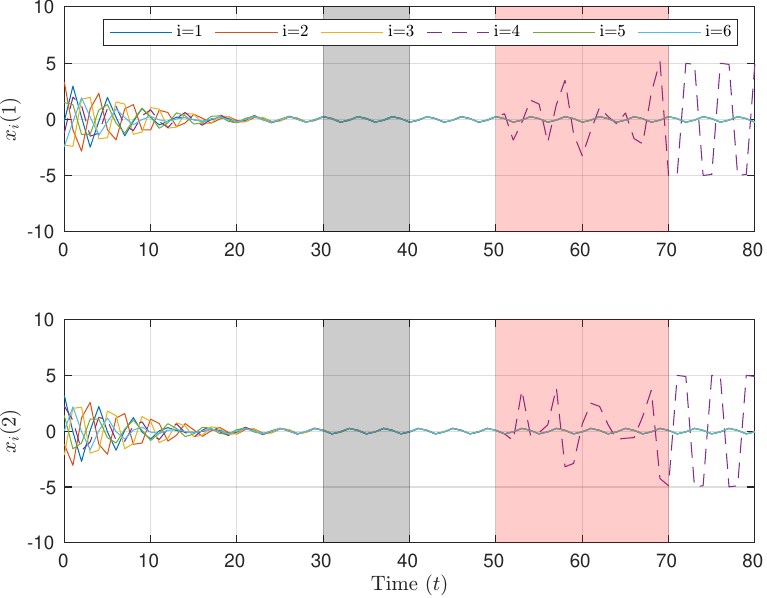}
\centering
\caption{The states $x_i$ of the MAS under $F=2$ locally adversarial attacks without the detection mechanism. The gray and red areas represent the periods under the adversarial link and agent, respectively.}
\label{whl9-fig:7}
\end{figure}

\begin{figure}[!ht]
\includegraphics[width=0.85\columnwidth]{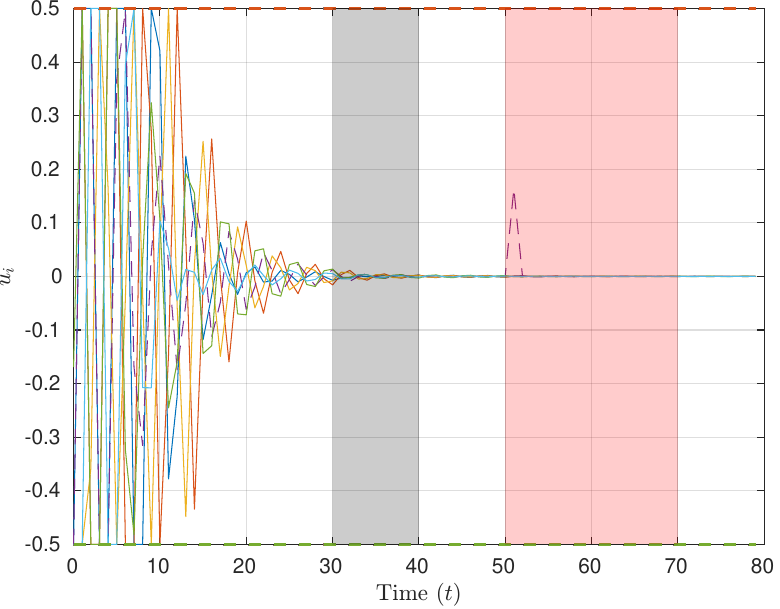}
\centering
\caption{The control inputs $u_i$ of the MAS under $F=2$ locally adversarial attacks with the detection mechanism. The gray and red areas represent the periods under the adversarial link and agent, respectively.}
\label{whl9-fig:8}
\end{figure}

In contrast, one can see from Fig. \ref{whl9-fig:7} that the resilient consensus of normal agents is achieved under \textbf{Algorithms}~\ref{whl9-alg:1} and \ref{whl9-alg:2}. This result is consistent with \textbf{Theorem}~\ref{whl9-thm:2}. Clearly, the control input constraints are satisfied as illustrated in Fig. \ref{whl9-fig:8}.

\textbf{Example 2: The resilient platoon control of connected and automated vehicles.}

We consider a group of four vehicles under $1$-local adversarial attacks, which move along a single lane with the desired distance gap $d=15$m and the same speed. Note that the adversarial attack is randomly generated from the interval $[-4,4]$ for $k\in\mathbb{N}_{[31,35]}$ and injected into the follower vehicle $3$. The longitudinal dynamics of vehicle $i$, $i=0,1,\dots,4$ is described by 
\begin{equation}\label{whl9-eq:34}
{x}_i(t+1)=Ax_i(t)+Bu_i(t),
\end{equation}
where $x_i(t)=[s_i(t), v_i(t),a_i(t)]^\T\in\mathbb{R}^3$ includes the position $s_i(t)$, the speed $v_i(t)$ and the acceleration $a_i(t)$; $$A=\begin{bmatrix}
1&T&0.5T^2\\
0&1&T\\
0&0&1-\frac{T}{\tau}\\
\end{bmatrix} \ \text{and}\  B=\begin{bmatrix}
0\\
0\\
\frac{T}{\tau}
\end{bmatrix},$$
with $T=0.2$ and $\tau=0.6$ being the sampling time interval and the vehicle engine constant, respectively. The control input constraints are $|u_i(t)|\leq3$. There are constraints on the vehicle system states, i.e., $0\text{m/s}\leq v_i(t)\leq30$m/s and $|a_i(t)|\leq3\text{m/s}^2$. Note that the virtual lead vehicle $0$ is set to run along 
\begin{equation}
  v_0(t) =
    \begin{cases}
      20\text{m/s}, &t\leq 6\text{s}, \\
      20+2.5t\text{m/s},& 6\text{s}<t\leq 10\text{s},\\
      30 & 10\text{s}<t\leq 16\text{s},
    \end{cases}       
\end{equation}
with $s_0(0)=0$ and $a_0(0)=0$. Each follower vehicle $i$, $i=1,2,3,4$ can receive the information from the lead vehicle $0$. 
The initial states of four agents are $x_1(0)=[-0.2, 20, -0.4]^\T$, $x_2(0)=[15.3, 20, -0.4]^\T$, $x_3(0)=[-30.9, 20,0]^\T$ and $x_4(0)=[-45.7, 20,0.1]^\T$, respectively. The initial network is configured as $\mathcal{N}_1(0)=\{2,4\}$, $\mathcal{N}_2(0)=\{1,3\}$, $\mathcal{N}_3(0)=\{2,4\}$ and $\mathcal{N}_4(0)=\{1,3\}$. 

The prediction horizon is $N=15$ and the estimation error set is $\Delta=\{x\in\mathbb{R}^3\mid\|x\|\leq0.5\}$. The weighting matrix is ${\Psi}=1$ and the pre-designed consensus gain matrix is designed as $K(0) = [-0.4042,-1.0015,-0.5387]$.

\begin{figure}[!ht]
\includegraphics[width=1\columnwidth]{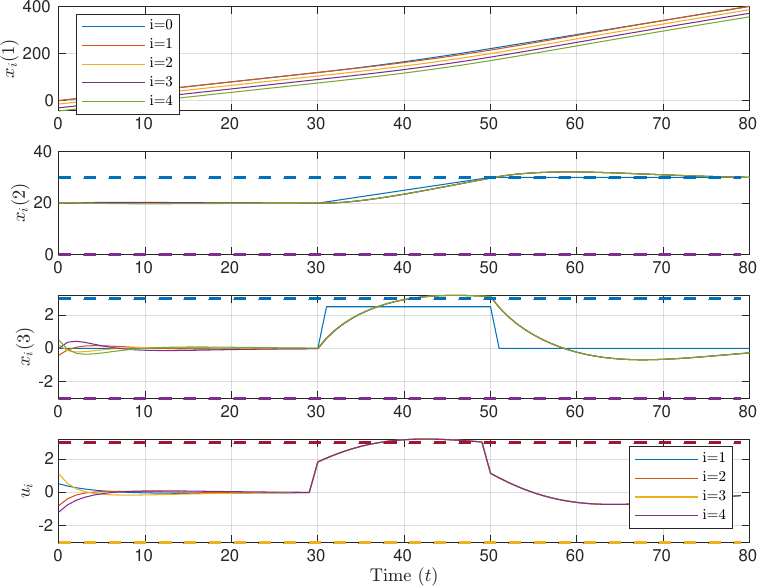}
\centering
\caption{The states $x_i$ and control inputs $u_i$ of the CAVs without adversarial attacks under the pre-designed consensus protocol.}
\label{whl9-fig:9}
\end{figure}

\begin{figure}[!ht]
\includegraphics[width=1\columnwidth]{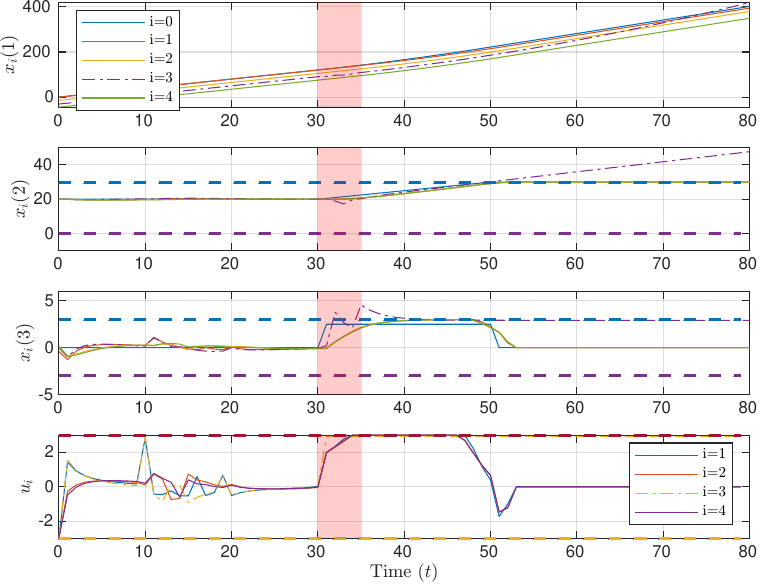}
\centering
\caption{The states $x_i$ and control inputs $u_i$ of the CAVs under $F=1$ local adversarial attacks with the detection mechanism. The red area represents the periods under the adversarial attack.}
\label{whl9-fig:10}
\end{figure}
Fig.~\ref{whl9-fig:9} and Fig.~\ref{whl9-fig:10} illustrate simulation results of the CAVs under the pre-designed consensus protocol and the proposed DMPC-based resilient consensus protocol, respectively. As shown in Fig. \ref{whl9-fig:9}, while the platoon control of the CAVs without adversarial attacks can be achieved by implementing the pre-designed consensus protocol, the constraints of the control input and system states cannot be guaranteed. 

The proposed DMPC-based resilient consensus protocol is implemented on the CAVs under $F=1$ local adversarial attack leading to the simulation results (see Fig. \ref{whl9-fig:10}). The normal vehicles (including vehicle $1,2,4$) are able to keep the desired distance while satisfying the control input and system state constraints. In contrast, the resilience of the CAVs under adversarial attacks cannot be ensured when the distributed attack detection mechanism is not implemented. These results are consistent with the theoretical results presented in this article.

\section{Conclusion}
\label{whl9-sec:7}
For the resilient consensus problem of the constrained MAS with adversarial attacks, we have proposed a novel DMPC-based consensus protocol, which integrated the pre-designed consensus protocol and the DMPC optimization. The optimal consensus protocol allowed the convergence analysis for the unconstrained MAS; it was updated when the cyber-attacks that caused the change of the communication networks occurred. The DMPC optimization was further introduced to handle the practical constraints while achieving the suboptimal consensus performance. Moreover, we developed a distributed attack detection algorithm to detect adversarial attacks, including Byzantine and malicious link/agent attacks. The proposed attack detection algorithm features the advantage that significantly relaxes the robustness requirement of the communication networks in contrast to the well-known MSR-type algorithms. For the general linear constrained MAS under attacks, we gave the sufficient conditions to ensure the proposed method's recursive feasibility and reach the resilient consensus. The effectiveness of the distributed detection mechanism was also analyzed for different types of attacks. Finally, the simulation results were provided to verify the effectiveness of the theoretical results. 

This work focuses on the resilient consensus problem of the homogeneous linear MAS over the undirected networks, which offers several potential avenues for the future research. Future works include extensions of the proposed method to 1) the constrained Lagrangian systems \cite{feng2019connectivity}; 2) heterogeneous MAS with limited communication resources \cite{ran2021practical}; and 3) the cases with uncertainties \cite{li2018robust}. In particular, we expect the proposed method to be further explored to achieve the resilient and privacy-preserving consensus \cite{ruan2019secure}.




\bibliographystyle{plain}        
\bibliography{whl9}           



\end{document}